# Statistical modeling: the three cultures

Adel Daoud[1,2,3*] and Devdatt Dubhashi[3]

1. Department of Sociology and Work Science, University of Gothenburg, Gothenburg, Swedenw
2. Center for Population and Development Studies, Harvard University Cambridge, The United States
3. The Data Science and AI division, Department of Computer Science and Engineering, Gothenburg, Sweden

* corresponding author: adaoud@hsph.harvard.edu

**Author contributions**: Conceived the research topic, led the research, and wrote the manuscript: AD. Commented and revised the manuscript: AD and DD.

**Acknolwedgemts:** We are grateful to the valuable feedback provided by Fredrik Johansson. All remaining errors and omissions are entirely our responsibility.

**Abstract**: Two decades ago, Leo Breiman (2001) identified two cultures for statistical modeling. The *data modeling culture* (DMC) refers roughly to practices aiming to conduct statistical inference, $\hat{\beta}$, on one or several quantities of interest. The *algorithmic modeling culture* (AMC) refers to practices defining a machine-learning (ML) procedure that generates accurate predictions, $\hat{Y}$, about an event of interest (outcome). Breiman argued that statisticians should give more attention to AMC than to DMC, because AMC provides novel procedures for prediction. While twenty years later, DMC has perhaps lost some of its dominant role in statistics because of the data-science revolution, we observe that this culture is still the *modus operandi*—the leading practice of a group—in the natural and social sciences. DMC is the modus operandi because of the influence of the established scientific method, called *the hypothetico-deductive scientific method*. Despite the incompatibilities of AMC with the hypothetico-deductive scientific method, among some research groups, AMC and DMC cultures mix intensely. We argue that this mixing has formed a fertile spawning pool for a mutated culture: a *hybrid modeling culture* (HMC) where prediction and inference have fused into new procedures where they reinforce one another. This article identifies key characteristics of HMC, thereby facilitating the scientific endeavor and fueling the evolution of statistical cultures towards better practices. By *better*, we mean increasingly reliable, valid, and efficient statistical practices in analyzing causal relationships. In combining inference and prediction, the result of HMC is that the distinction between $\hat{Y}$ and $\hat{\beta}$—taken to its limit—melts away. We qualify our "melting-away" argument by describing three HMC practices, where each practice captures an aspect of the scientific cycle: ML for causal inference, ML for data acquisition, and ML for theory prediction.

# 1 Introduction

Two decades ago, Leo Breiman (2001) identified two cultures for statistical modeling. The *data modeling culture* (DMC) refers roughly to practices aiming to conduct statistical inference on one or several quantities of interest. By *statistical inference*, we mean a procedure tailored to estimate a quantity $\hat{\beta}$ such that the difference between the true quantity $\beta$ is as small as possible. This true quantity $\beta$ is assumed to exist independently of the statistical model producing $\hat{\beta}$. For example, at any particular time, Harvard freshman have a true average population height $\beta_h$, yet a procedure estimating this height, $\hat{\beta}_{height}$, must account for a variety of disturbances related to sampling and measurement tools. Similar disturbances exist in causal inference (fully defined in Section 3.1), which is a particular type of statistical inference (Imbens and Rubin 2015; Pearl 2009). For example, doctors estimating the treatment (causal) effect, $\hat{\beta}_{cure}$, of a new cure on their patients assume that this cure has a true effect $\beta_{cure}$, but this estimation is hampered by, among other things, the characteristics of the patients and how the cure is administered. A procedure is unbiased when the difference $\beta - \hat{\beta}$ is negligible in expectation. Breiman argued that DMC is the dominant mode of operations in statistics. While twenty years later, this culture has perhaps lost some of its dominant role in statistics because of the data-science revolution, we observe that this culture is still the *modus operandi*—the leading practice of a group—in the natural and social sciences.

DMC is the modus operandi because of the influence of the established scientific method, called *the hypothetico-deductive scientific method* (Hempel 1965; Popper 2002). This scientific method consists of cycles of deductively formulating a hypothesis from substantive theory, testing the hypothesis in a model and against data, and revising theory. Deductive reasoning uses universal propositions (hypotheses derived from general theories) to explain specific events. By *explanation*, we mean a theory that demonstrates how two or more events are causally related (Bhaskar 2008). When enough evidence has been collected that challenges existing truths, an entire paradigm can fall in favor of a new one (Kuhn 2012). This scientific method favors DMC over other modeling cultures because DMC enables both descriptive and causal thinking.

The *algorithmic modeling culture* (AMC) refers to practices defining a procedure, *f*, that generates accurate predictions, $\hat{Y}$, about an event (outcome), $Y$ (Breiman 2001). By *accurate*, we mean predictions that are as similar as possible to the true event that *f* has yet not encountered (Hastie, Tibshirani, and Friedman 2009). The smaller the difference $Y - \hat{Y}$, the higher the similarity. We define a *procedure* as an algorithm, or a function, that takes some input $X$, operates on this input $f(X)$, and then, produces an output $f(X) = \hat{Y}$. Often, this procedure is defined inductively, that is, by following the patterns in the data (Hastie et al. 2009). While working in AMC, scholars care about the statistical interpretability of the used procedure only so far that it furthers their pursuit to generate accurate predictions (Lipton 2017). To identify causal relationships between *X* and *Y* is a peripheral question, if at all interesting. This culture is the modus operandi of many strands in engineering, computer science, the industry, and policy (Sanders 2019). As AMC procedures do not align with the hypothetico-deductive scientific method (Molina and Garip 2019; Mullainathan and Spiess 2017; Turco and Zuckerman 2017; Watts et al. 2018), this culture has not taken root in the applied sciences beyond these strands.

Despite the incompatibilities of AMC with the hypothetico-deductive scientific method, among some research groups, AMC and DMC cultures mix intensely. We argue that this mixing has formed a fertile spawning pool for a mutated culture: a *hybrid modeling culture* (HMC) where prediction and inference have fused into new procedures where they reinforce one another. As Section 3 discusses, scholars use these procedures in the pursuit of explaining how two events are causally connected by blending $\hat{Y}$-prediction problems and $\hat{\beta}$-inference problems to the point that it is difficult to tell them apart (Molina and Garip 2019; Mullainathan and Spiess 2017; Yarkoni and Westfall 2017). These fused procedures are still compatible with the hypothetico-deductive scientific method but stretch beyond it because they allow for a large portion of inductive reasoning. Such reasoning infers universal claims from particular observations. While this hybrid culture does not occupy the default mode of scientific practices, we argue that it offers an intriguing novel path for applied sciences.

This article aims to identify key characteristics of what we named HMC, thereby facilitating the scientific endeavor and fueling the evolution of statistical cultures towards better practices. By *better*, we mean increasingly reliable, valid, and efficient practices in analyzing causal relationships. We execute our argument "by examples," meaning that we will selectively review trends in applied research as proof for the existence of HMC. From these examples, we will pinpoint the defining characteristics of HMC.

Before discussing HMC, we identify two trends that nourish the emergence of it. First, to a considerable extent, HMC has emerged from applied computational research that closely interfaces with statistics and computer science—that is, data science (Efron and Hastie 2016). For example, this close interfacing in the social sciences is known as computational social science, which denotes any scientific study that develops or uses computational methods to typically large-scale and complex social and behavioral data (Keuschnigg, Lovsjö, and Hedström 2017; Lazer et al. 2020). The stream of new data sources—administrative data, content, and networks of social media, digitalized corpora, video, audio—explain the relevance of such a computational approach (Jordan and Mitchell 2015). Similar computational approaches exist under the brands of digital humanities (Gold 2012), computational psychology (Sun 2008), computational economics (Tesfatsion and Judd 2006), computational epidemiology (Marathe and Vullikanti 2013; Salathé et al. 2012), computational biology (Noble 2002), to mention a few. All these computational approaches emerged by the beginning of the twenty-first century, and they provide clues to why DMC or AMC is insufficient alone to cover the new demands of the scientific endeavor: that is, to provide systematic explanations of events of reality, and thereby to deepen our knowledge of them (Bhaskar 2008).

A second feeding ground for the evolution of HMC is the causal-inference revolution. While a randomized control trial (RCT) remains the safest way to rinse out the contaminating effect of confounding and the least assumption-demanding method to identify causality (Fisher 1935), scholars face ethical and practical limitations when applying a RCT (Deaton and Cartwright 2016). For example, to estimate the causal impact of events such as economic crises (Daoud et al. 2017), famines (Daoud 2017; Sen 1981), or natural disasters (Daoud, Halleröd, and Guha-Sapir 2016), on children's well-being, scholars would need to administer such events to a treatment and control group of children. Ethical and practical limitations of RCT are two sources fueling the causal revolution that has resulted in a myriad of new approaches for inferring causality in observational data (Angrist and Pischke 2014; Hedström and Manzo 2015; Hernan and Robins 2020; Imai 2018; Imbens and Rubin 2015; King 1998; van der Laan and Rose 2011; Morgan and Winship 2014; Pearl and Mackenzie 2018; Peters, Janzing, and Schölkopf 2017; VanderWeele 2015). As this revolution has partly evolved from computer science, partly from biostatistics, statistics, and economics, scholars have creatively combined tools from DMC and AMC. As HMC synthesizes the strengths of DMC and AMC, this synthesis has resulted in practices better adapted to twenty-first-century requirements than what DMC or AMC alone can offer.

The remainder of our argument is less concerned with why HMC has emerged and more with characterizing it.

## 2 Thinking predictively and inferentially

Scholars in the applied sciences think and operate mainly through the hypothetico-deductive scientific method (Danermark et al. 2002; Kuhn 2012). This method is a philosophy of science, defining how scientific inquiry should be conducted (Hempel 1965; Popper 2002). Using substantive theories, scholars articulate their causal and descriptive knowledge in falsifiable hypotheses and operationalize them in data. Thus, this method is "hypothetico." Hypotheses are postulated by specifying a statistical model representing how the data (events) are generated. Thus, this method is "deductive." The method aims to test whether the assumed data-generating process of the model matches the collected data (King 1998). If the null hypothesis—that two events are statistically unrelated—receives no support in data, this hypothesis has been falsified under the stipulated model. If falsified, then the alternative hypothesis—that two events are related—receives support. Based on the modeling results, scholars revise their knowledge. And so, the circle continues. The requirement of testing substantive theories in an interpretable statistical model is one of the appeals and endorsement of DMC in the applied sciences (Breiman 2001).

To compare DMC and AMC—and later characterize HMC—we define the following terminology. Scholars formulate, test, and develop substantive theories, $T_1, \ldots, T_k$, about a causal system. A *causal system* is a set of events and relationships between events in a domain of interest.[1] Because elements of the causal system rarely reveal themselves directly to the human senses, scholars theorize about the existence of events and their causal relationships. By *theory*, $T_k$, we mean a set of concepts that enable formulating descriptions, predictions, hypotheses, or explanations about events populating a causal system (Swedberg 2017). Each $T_k$ map into a directed acylic graph (DAG), $G$, that formalizes and visualizes a potential manifestation of the causal system of interest (Pearl 2009).[2] While two or more theories often compete for proposing the best explanation—meaning how well they account for observed data—they do not have to be mutually exclusive. Scholars can formulate theories at various abstraction levels, but to get tested empirically, they need to match what is measurable (Merton 1968).

Figure 1 shows the progression of knowledge under DMC with two stylized DAGs, $G_1, G_2$ and $G_3$ competing to explain famines. A significant debate raged between Malthusians and Senians on whether food scarcity is a necessary event to cause famines (Daoud 2007; Devereux 2007; Sen 1981)—a debate that still influences ecology and sustainability research (Daoud 2018; Dobkowski and Wallimann 2002; Meadows et al. 1972). Three centuries ago, Thomas Malthus argued that while population size increases geometrically, food supply increases arithmetically (Malthus 1826). Because population size will outstrip food supply, famines will eventually emerge to balance their relationship. Based on Malthus' theory $T_1$, DAG $G_1$ depicts this famine-causal system. Amartya Sen challenged this explanation by showing that famines—at least in the modern era—can arise even when there is sufficient or abundant food (Sen 1981). Especially when social inequality is high, vulnerable groups run a higher risk of unemployment than other groups. Unemployment causes a loss of income and individuals' capability to purchase food. This loss of capability—what Sen named entitlement failure—results in starvation. In Sen's theory $T_2$, the DAG $G_2$ is a better representation of the causal system of famines. Although Sen acknowledged that population size and food-supply shortage can cause famines, such shortage is "…one of many possible causes," during the last century (Sen 1981:1). Thus, Sen's theory argues that population size and food supply have roughly no effect on the probability of famines, that is, $\beta_3, \beta_2 \approx 0$.

---
[1] For causal inference, we use notations from both *do*-calculus (Pearl 2009) and the potential outcomes framework (Imbens and Rubin 2015).
[2] Systems that do not map into a well-defined causal order, may map to other graphical representations (e.g., sequence, equilibrium, or transition graphs).

Subsequent theoretical development $T_3$ use both theories to offer an even more robust approach to explaining events of famines (Daoud 2007, 2010, 2017). A critical conceptual movement is to disentangle societal and individual-level starvation. As $G_3$ shows, while Malthus theory explains when famines are likely to arise at the societal level, it cannot explain which individuals will like to starve to death. Sen's theory identifies these individuals by their failing entitlements to food (Reddy and Daoud 2020). Additionally, the entitlements-causal path ($\beta_4$ via $\beta_6$ to $\beta_7$) shows that famines can arise even if there is no societal scarcity.

As exemplified in the research progress of famine research, DMC practices encourage scholars to pursue ever deeper knowledge production. Scholars aim to quantify, describe, and evaluate key events and their relationships. A *description* is a statistical inference about aspects of the distribution of one or more events, represented as a node in a DAG). For example, let FAMINES be a binary random variable measuring the occurrence of famines in a well-defined time (e.g., in years 1900 to 2000) and space (e.g., Europe, Asia, or Africa). Then, the expectation $E[\text{FAMINES}]$ captures the average occurrence of famines in that time and space. Small caps denote variables in a DAG. A description of two or more events is associations, meaning that their correlation may or may not be causal.

Often scientific debates refer to the content of a causal system. For example, scholars debate what events populate such a system and how these events cause and effect each other. This content also defines the conditions under which empirical associations may be interpreted as causal or confounded. As Section 3.2 discusses, a confounded association is an association between two events $W$ and $Y$ that is partly determined by a common cause, $C$. If an association is confounded, then the association between $W$ and $Y$ is biased (Hernan and Robins 2020). To statistically capture causal associations, scientists analyze the joint (association) distribution of the events of interest (e.g., $p(Y, W, C)$) and how this joint distribution factorizes. In a causal system, a factorization defines the causal order statistically among events. For example, the distribution $p(\text{FAMINES}, \text{SCARCITY}, \text{POPULATION}, \text{FOODSUPPLY}, \text{ENTITLEMENTS})$ have many potential factorizations depending on substantive theory. While in Malthus' $G_1$ stipulates the following factorization,

$$p(\text{FAMINES}, \text{SCARCITY}, \text{POPULATION}, \text{FOODSUPPLY}, \text{ENTITLEMENTS}) = \\ p(\text{FAMINES}|\text{SCARCITY})\, p(\text{SCARCITY}|\text{POPULATION}, \text{FOODSUPPLY}) p(\text{ENTITLEMENTS}) \\ p(\text{FOODSUPPLY}) p(\text{POPULATION}),$$

Sen's $G_2$ declares that the following factorization is the best approximation of famines,

$$(\text{FAMINES}, \text{SCARCITY}, \text{POPULATION}, \text{FOODSUPPLY}, \text{ENTITLEMENTS}) = \\ p(\text{FAMINES}|\text{SCARCITY})\, p(\text{SCARCITY}|\text{ENTITLEMENTS}) p(\text{ENTITLEMENTS}) \\ p(\text{FOODSUPPLY}) p(\text{POPULATION}).$$

The scientific and policy differences are large, depending on whether $G_1$ or $G_2$ is the best representation of reality (Daoud 2017). If $G_1$ best represents the causes and effects of a famine, then policymakers should produce more food to counter famines; conversely, if $G_2$ is better, then this theory stipulates that policymakers should reduce social inequality to reduce the probability of famines (Daoud 2015a; Halleröd et al. 2013). Such mixing of causal reasoning and ethics is an emerging field in computer science (Daoud, Herlitz, and Subramanian 2020; Kusner et al. 2017).

In DMC, to estimate the causes and effects of famines, scholars collect famine data and test their stipulated models. The model that fits the sample best, receives scientific support. Scientific debates tend to amplify when different samples yield support to different models. Scholars evaluate support for a statistical model by interpreting how different factorizations match the sample. A factorization can be estimated with a variety of statistical models $f$, yet to retain interpretability, DMC-influenced scholars often use a model from the linear-function class family, $f \in F$. A linear model of $G_1$ could then have the following stylized statistical form, $\text{FAMINES}_i = f_1(\text{SCARCITY}_i) + e_i$, where $f_1(\text{SCARCITY}_i) = c_0 + \beta_1 \text{SCARCITY}_i$ and $i$ indexes instances of famine. The first equality follows from $G_1$, and the second equality follows from the assumed linear model. In this example, a hypothesis operationalizes an aspect of $T_1$ and its model evaluates if $\beta_1$ is different from zero. Generally, a *hypothesis* operationalizes an aspect of $T_k$ stipulating the existence of a causal relationship (edges) between events (nodes). Scholars use interpretable statistical models so that they can imprint their hypothesis into these models. This is why linear models are popular because they make this imprinting straightforward (Hastie et al. 2009).

The goal of testing theories about causal systems through interpretable statistical models explains why a DMC-operating scholar would turn suspicious of throwing a set of predictors into a machine-learning (ML) model. ML is the sub-discipline of computer science that studies how models—or algorithms—can learn from data (Efron and Hastie 2016; Hastie et al. 2009). As many ML models are nonparametric, it is unclear how to unpack and interpret them as one would do for parametric models (Lipton 2017). Without a clear theoretical rationale for using ML, DMC scholars see little value for using them in the scientific process (Breiman 2001).

Conversely, ML models lie at the heart of many AMC practices. A key assumption of AMC is that a system produces a set of associations between a set of inputs, $X$, and outputs, $Y$. The relationship between $X$ and $Y$ may or may not be causal. The overarching goal is to develop an algorithm $f$ that operates on these inputs and produce the best possible predictions $\widehat{Y}$ of $Y$ that $f$ has not been observed yet. Figure 2 shows a stylized graphical representation of the system of associations among inputs and outputs. This graph is still directed but not necessarily representing a causal relationship, and thus, we call it a Bayesian network (Pearl 2009). All DAGs are Bayesian networks, but not all Bayesian networks are DAGs.[3] In a Bayesian network, an edge denotes an association between two nodes with no causal interpretation. In the graph of Figure 2, all inputs $X_k$ are connected because the association is assumed to flow in all directions. A linear model and a fully connected graph has $\beta_1, \dots, \beta_k$ association between $Y$ and each $X_k$, and $\beta_{k+1}, \dots, \beta_{\binom{k}{2}}$ pairs of association among $X_k$. Because scholars evaluate the predictive performance of $f$ comparing $\widehat{Y}$ using a held-out set $Y$, scholars pursue to interpret these association only as a subordinate priority (Doshi-Velez and Kim 2017)—if at all.

---

[3] Undirected graphs are called Markov networks (Pearl 2009).

As the main goal of AMC is not to develop causal knowledge, $T_k$, about a system, AMC demotes causal reasoning (Pearl and Mackenzie 2018). Despite the lack of causal reasoning, AMC exhibits many ML innovations across several domains. In robotics, autonomous vehicles are capable of driving far distances without human supervision; in the arts, the Next Rembrandt project has shown how image-recognition algorithms can recreate a Rembrandt painting at the level of mastery that humans have difficulty telling them apart from an original Rembrandt;[4] similarly, in music, MuseNet composes symphonies at par with Mozart, Chopin, or Beethoven;[5] in linguistics, OpenAI's Generative Pre-trained Transformer 3 (GPT-3) writes articles, songs, and manuscripts with impressive coherency; in gaming, DeepMind's algorithms AlphaStar and AlphaGo compete at the grandmaster level in the computer game Starcraft II (Vinyals et al. 2019) and the board game Go (Silver et al. 2017), respectively. In our famine example, ML is used to build early-warning systems. The Famine Action Mechanism—a collaboration among the World Bank, United Nations, Food and Agricultural Organizations, and others—and the Famine Early-Warning-System-Network[6] (led by USAID) are concrete policy examples tailored to minimize the outbreak of famines.

Although much remains to be proven before the same algorithm—strong artificial intelligence—may roam across all these domains, AMC innovations are noteworthy because scholars have developed each algorithm without explicitly knowing the causal connections among the events in the system (Domingos 2015). ML models learned the relevant associations inductively from data.

The advancements of AMC resonate with Karl Pearson's idea that statistical correlation—predictability—between *X* and *Y* is what scholars should search for to advance science. For Pearson, correlation *is* causation. He argues,

> Take any two measurable classes of things in the universe of perceptions, physical, organic, social or economic, and it is such a dot or scatter diagram, which we reach with extended observations. In some cases the dots are scattered all over the paper, there is no association of A and B ; in other cases there is a broad belt, there is only moderate relationship; then the dots narrow down to a "comet's tail," and we have close association. Yet the whole series of diagrams is continuous; nowhere can you draw a distinction and say here correlation ceases and causation begins. Causation is solely the conceptual limit to correlation, when the band gets so attenuated, that it looks like a curve. (Pearson 1911:170)

Despite the absence of causal reasoning, scholars have produced many AI innovations using the principles of AMC. However, we argue that this absence is a major limitation for the advancement of scientific knowledge (Darwiche 2017; Pearl and Mackenzie 2018). Developing ML that paints like Rembrandt and composes symphonies like Mozart, but not knowing exactly what causes what in their respective machinery, is not sufficient to advance scientific knowledge. Similarly, training an ML-powered early-warning system of famines works will support policymakers in countering starvation. Yet if scholars are unable to unpack what affects what in such a system, then the science of famines gains little from such ML-powered systems (Rudin 2019). In his commentary, Brad Efron points at a similar limitation by arguing that Breiman overstates the role of prediction and that AMC does not assist in "the identification of causal factors." (p. 219). Breiman acknowledges the absence of causal reasoning in his response to Efron's commentary, but he maintains his emphasis on predictions,

> I agree that often '..statistical surveys have the identification of casual factors as their ultimate role.' I would add that the more predictively accurate the model is, the more faith can be put into the variables that it fingers as important. (Breiman 2001:229)

---

[4] nextrembrandt.com
[5] openai.com/blog/musenet/
[6] www.fews.net

Even if DMC suits the scientific method better, it suffers from at least two limitations. First, because of the suspicion towards AMC-predictions, DMC scholars rely on analog methods to collect information about the causal systems (Salganik 2017). Surveys, experiments, and interviews are standard methods as they give full (human) control over the data collection process. However, in the digital age, these analog approaches limit the speed and type of data collected. Second, in agreement with Breiman (2001), we argue that a more severe limitation of DMC is that it relies mainly on model validation. A scholar formulates a statistical model and then tests that model against data. Using various goodness-of-fit metrics, this scholar then draws a set of conclusions. Yet, these conclusions are often more telling about the assumed model's structure and less about the causal system. If the statistical model is a poor representation of this causal system—for example, that the relation between $X$ and $Y$ is not linear—these conclusions may be misleading despite a relatively better fit of one model over another.

Table 1 summarizes the key characteristics of DMC and AMC (Breiman 2001). Although famines exemplify our argument, these characteristics generalize to any scientific domain. The goal of DMC is to identify the causal quantity $\hat{\beta}$ of the relationship between food availability and famine. In AMC, the goal has shifted to predicting famines, $\hat{Y}$, from any relevant input data, $X$ (Mullainathan and Spiess 2017). While the first stipulates untestable statistical assumptions, the second relies on black-box models. Because different disciplines have made major breakthroughs under the influence of AMC and DMC, a warranted question is then: to what extent can they be synthesized to further the scientific endeavor.

Table 1 [about here]

## 3 A hybrid statistical-modeling culture: a unifying framework fueling the evolution of scientific practices

By upholding disciplinary traditions, university departments also inadvertently create cultural silos where DMC and AMC practices dwell (Lazer et al. 2020). Changing these traditions is challenging. Nonetheless, in some places—such as the Harvard Data Science Initiative, Harvard Institute for Quantitative Social Science, Chalmers AI Research Centre (Sweden), Institute for Analytical Sociology (Sweden), the Alan Turing Institute (U.K.)—where the isolating effect of departmental silos have been dismantled, a new hybrid statistical-modeling (HMC) culture is emerging.

As HMC evolutionary descendants from DMC and AMC, it benefits by copying the useful elements from each of the two cultures and mutating them into new practices. Copying the aim of DMC and submitting to the hypothetico-deductive scientific method, HMC still regards the overarching aim of science comprising to identify and explain the causal link between $X$ and $Y$ (Pearl and Mackenzie 2018). This aim manifests in the advancement of substantive theories, explanations, and hypotheses that cyclically gets tested with statistical models and against data. Nonetheless, instead of relying on the commonly used statistical models in DMC, HMC leverages the arsenal of ML algorithms developed under AMC, thereby increasing scholars' modeling power (Kuang et al. 2020; Künzel et al. 2018; van der Laan and Rose 2011, 2018; Peters et al. 2017). In combining inference and prediction, the result of HMC is that the distinction between $\hat{Y}$ and $\hat{\beta}$—taken to its limit—melts away.

We qualify our "melting-away" argument by describing three HMC practices, where each practice captures an aspect of the scientific cycle. Table 2 shows what these three practices constitute: ML for causal inference, ML for data acquisition, and ML for theory prediction. Although these three goals exist partly in DMC (i.e., inference) and AMC (i.e., prediction), HMC fulfils them by blending inferential and predictive thinking. For example, one DMC goal comprises estimating causal effects by deductively inferring the statistical relationship between *W* and *Y*, under an assumed DAG. This goal exists in HMC as well. Nonetheless, while deductive causal evaluation agrees with one goal of HMC, there is no DMC equivalent for letting an algorithm to inductively discover how this DAG, including *W* and *Y*, should be ordered to begin with (Glymour, Zhang, and Spirtes 2019; Peters et al. 2017). Training such algorithms is a HMC specific problem, called *causal discovery*. As the next section shows, such causal-discovery algorithms—like all the other HMC practices we will discuss—combine ML prediction and inference to such a high degree that neither DMC nor AMC can comfortably host them.

The next section exemplifies the stylized characteristics of HMC by showing how specific statistical practices use prediction and inference in tandem. While that section focuses on causal inference, HMC practices generalize to statistical inference in general. For example, while the goal of ML for causal inference is to evaluate the causal relationship between two events (e.g. the impact of food scarcity on the probability of a famine event), the goal of ML for data acquisition is to infer (measure) the value of these events separately (e.g., the mean and variance of famines and food supply in the Bengal 1940s).

Table 2 [about here]
f

## 3.1 ML for causal inference

Before describing how ML aids in inferring causality in HMC (Table 2 column one), we will refine our definition of what we mean by causal inference. This definition applies to both DMC and HMC. Table 3 illustrates an observed data matrix of four individuals with fictitious variable values.

Table 3 [about here]

We define a cause of interest as a binary variable, $W$. Instead of merely recording each individual's outcome as observed by the data, $Y_i$, we assume that each individual *i* has two potential outcomes (Imbens and Rubin 2015). One potential outcome represents the outcome when the individual takes the treatment, $Y_i^1$, and one where they do not take it, $Y_i^0$. The causal effect for each individual *i* is then the difference between these two potential outcomes:

$$\tau_i = Y_i^1 - Y_i^0$$

If we could observe both potential outcomes, we could then directly compute $\tau_i$ and thus identify individual-level causal effects. However, the observed outcome—as supplied by the data—is a function of both the treatment and the two potential outcomes, $Y_i = (W-1)Y_i^1 + WY_i^0$. This function shows that the observed data, exemplified in Table 3, reveals only one of these two potential outcomes, yet both are required to identify causal effects. This impossibility of observing both potential outcomes is known as the *fundamental problem of causal inference*. Much of the causal-method development pertains to reasoning about identifiability and defining procedures for calculating causal effects from observational data (Hernan and Robins 2020; Imbens and Rubin 2015; Pearl 2009; Peters et al. 2017). *Identifiability* means articulating a set of assumptions that allow a model to calculate a causal effect from data.

Vibrant literature in the overlap between computer science, econometrics, and statistics combine ML and causal methodology to identify causal effects (Angrist and Pischke 2014; Athey and Imbens 2017; Athey, Tibshirani, and Wager 2019; Chernozhukov et al. 2018; Hedström and Manzo 2015; Hernan and Robins 2020; Hill 2011; Hirshberg and Zubizarreta 2017; Imai 2018; van der Laan and Rose 2011; Morgan and Winship 2014; Pearl and Mackenzie 2018; Peters et al. 2017; Sekhon 2009; VanderWeele 2015). A reoccurring theme in this overlap is the many creative combinations of methods where predictive AMC-type algorithms are used in DMC-type of causal inference. These methods summarize into at least four types of combinations.

In the first combination, scholars use ML to impute (predict) potential outcomes. As observed data only reveal one-half of the potential outcomes, the other half is regarded as missing data. One way of handling this fundamental problem is to cast it as a missing-data problem and proceed to identify conditions for imputing these data to populate all the $Y_i^1$ and $Y_i^0$ cells, based on the similarity of covariates $X$. These imputation procedures rely on common identifiability assumptions. One such central assumption is conditional independence (also known as conditional ignorability and conditional exchangability), $Y_i^1, Y_i^0 \perp W|X$. This mathematical statement means that the treatment is as-if randomly assigned conditional on one or more covariates.

Because ML excels in prediction tasks compared to commonly used parametric models, HMC-influenced scholars have developed many different procedures to predict potential outcomes (Künzel et al. 2018). For example, the *T-learner*—"T" stands for "two"—procedure defines one ML-algorithm $f_{w=1}(x_i) = E[Y = y_i|W = 1, X = x_i]$ trained on the treated group and another algorithm $f_{w=0}(x_i) = E[Y = y_i|W = 0, X = x_i]$ trained on the control group. Depending on the scientific problem, the scholar defines the type of algorithm—a Lasso, neural network, a random forest, or a collection of algorithms (an ensemble). After training, $f_{w=1}$ imputes potential outcomes for the control group and $f_{w=0}$ imputes these outcomes for the treated group. Based on the toy data of Table 3, $f_{w=1}$ trains on the variables of Jane and John, and imputes $Y_i^1$ for Joe and Jan; likewise, $f_{w=0}$ trains on Joe and Jan, and imputes $Y_i^0$ for Jane and John. This procedure culminates by calculating the difference $\hat{\tau}_i = \hat{Y}_i^1 - Y_i^0$ for the control group and $\hat{\tau}_i = Y_i^1 - \hat{Y}_i^0$ for the treated group, and then averaging over all groups $\hat{\tau} = E[E[\hat{\tau}_i|W = w_i]]$ to calculate the average causal effect.

The T-learner algorithm is one of several, but common to most of these ML algorithms is the procedure of imputing potential outcomes (Künzel et al. 2018) or imputing the treatment effect directly (Athey et al. 2019; Nie and Wager 2017). Consequently, in these sorts of HMC practice, $\hat{\tau}$-problems of HMC subsume $\hat{\beta}$-problems of DMC. As Section 2 discusses, a scholar under the influence of DMC would most likely articulate a set of assumptions for when the causal effect is identified (e.g., in a DAG) and specify a linear model to estimate this effect. For example, in our famine example depicted in Figure 1, a Malthusian scholar would argue that SCARCITY$_i$ (of food) causes FAMINES$_i$. This scholar may show the appropriateness of their assumption in the DAG, $G_1$, and proceed to specify the following stylized statistical model, FAMINES$_i = c_0 + \beta_1$SCARCITY$_i$, where $i$ indexes famine events. This model imprints the causal effect in the parameter $\beta_1$. If the true relationship between famines and scarcity is linear, this statistical model will capture the desired causal effect by interpolating between famine cases where scarcity was observed and where it was not. However, in most scientific domains, a scholar's preference to use a linear model follows more the desire to readily and transparently interpret a statistical model rather than following this scholar's knowledge that the complexities of reality are truly linear (Abbott 1988; Lipton 2017). Despite that linear models can capture nonlinearities via a variety of transformations, to model a nonlinear reality, ML for causal inference offer a more robust alternative (van der Laan and Rose 2018).

If a famine scholar followed the statistical practices of HMC instead of DMC, this scholar would formulate the same causal goal as a $\hat{\tau}$-problems, and now use an ML algorithm to estimate the causal effect by imputing potential outcomes (Künzel et al. 2018). Using the same DAG $G_1$, the scholar would, for example, use a T-learner to impute the probability of famine in cases where scarcity is present and where scarcity is absent. Although both $\hat{\tau}$ and $\hat{\beta}$ quantify the same causal effect, their key difference resides in that $\hat{\beta}$ refers mainly to a parametric setting whereas $\hat{\tau}$ refers to a nonparametric setting—a statistical-modeling nomenclature. To calculate $\hat{\tau}$, the scholar mobilizes the algorithmic power of AMC that is originally tailored for $\hat{Y}$-problems, but now recalibrated for $\hat{\tau}$-problems. By predicting (imputing) potential outcomes, the original distinction between $\hat{\beta}$ and $\hat{Y}$ has dissipated—melted away.

Predicting potential outcomes constitutes one necessary step in calculating counterfactuals, and thus, individual-level causal effects (Pearl 2009). The definition of potential outcomes and counterfactuals are closely related, but they refer to different scenarios. Potential outcomes refer to a scenario where the treatment assignment has not been made yet, and thus, *before* the treatment has been assigned, an individual has two potential outcomes $Y_i^1$ and $Y_i^0$. Counterfactuals refer to a scenario where the treatment has been assigned, but the scholar imagines what the outcome would have been having the treatment assignment been different. A counterfactual exists *after* the treatment has been assigned. For example, if the individual was assigned the treatment, and therefore, his or her factual outcome equals the potential outcome under treatment $Y = Y_i^1$, then this individual's counterfactual outcome is $Y_i^0$. Thus, counterfactuals enables retrospective reasoning of the form, "What if I had acted differently, would the result turned out the same" (Pearl 2019).

To calculate the value of a counterfactual, scholars must make assumptions about the noise (error) variables, $U = u$, in a causal system (DAG) (Pearl 2009). These variables represent any exogenous events—occurrences that are only indirectly relevant to the causes and effects of a DAG—that induce variations across individuals, and thus when this noise is known, it uniquely determines everyone's values in the data. These variations represent all factors particular to everyone, yet they are not necessary to the DAG, and thus, they are not explicitly specified in a DAG. For example, although an individual's genetics calibrate this person's physiology and thus nutritional intake, in famine situations, genetics do not directly add to explaining famine outcomes. Although in large samples, these variations cancel each other out when calculating the average treatment effect, $\tau = E[Y_i^1 - Y_i^0]$, these variations are key to calculate individual-level treatment effect, $\tau_i = Y_i^1 - Y_i^0$, for individual $i$.

As the context $X$ and these variations $U$ jointly determine the exact conditions when the individual toke a treatment versus did not take this treatment, we need to know both $X$ and $U$ to calculate $\tau_i$. A scholar can only know these quantities when the DAG and its structural causal relationships are known (Pearl 2009). When these are known, counterfactuals enable probability expression such as $P(Y = y^x | W = w', Y = y')$, which stand for "the probability of observing $Y = y$ had $W$ been $w$, given that we factually observed $W = w'$ and $Y = y'$. For example, in our famine case, his probability can refer to a specific Bengali farmer: would the farmer have survived $Y = y_i^1$ had the Bengali government distributed food coupons (entitlements to food) to farmers $W = 1$, given that this farmer starved to death $Y = y_i^0$ and did not receive coupons $W = 0$ (Daoud 2017, 2018). Despite that counterfactuals necessarily rely on stronger assumptions than calculating average effects, they present an exciting path for applied domains, from personalized medicine (Gottesman et al. 2019) to precision agriculture (Bauer et al. 2019), to incorporate retrospective thinking. Pearl takes this statement one step further by arguing that "Advances in graphical and structural models have made counterfactuals computationally manageable and thus rendered causal reasoning a viable component in support of strong AI" (Pearl 2019:1).

Predicting counterfactuals and potential outcomes enable an inductive analysis of treatment heterogeneity (Athey et al. 2019; Imai and Ratkovic 2013; Künzel et al. 2018). While average treatment effects capture the aggregated effect of an exposure on a population, treatment heterogeneity captures the group-specific effects disaggregated by sub-populations. For example, despite that a famine or an economic crisis is likely to affect an entire country adversely, some combination of socio-economic factors may protect certain groups better than others (Daoud and Johansson 2020). Traditional statistical approaches capture such treatment heterogeneity with the help of interaction models. Because these models are parametric, they often need a scholar to specify the product terms explicitly. If scholars hypothesized that ethnicity moderated the effect of entitlements in explaining famines, then they would deductively specify the following model, $\text{FAMINES}_j = c_0 + \beta_1 \text{ENTITLEMENTS}_j + \beta_2 \text{ETHNICITY}_j + \beta_3 \text{ENTITLEMENTS}_j \cdot \text{ETHNICITY}_j$, where $j$ indexes individuals. If the parameter $\beta_3$ is statistically significant, then that is evidence for treatment heterogeneity.

However, an ML model for treatment heterogeneity does not require such explicit specification. In this model, the treatment heterogeneity for our famine example is defined as the conditional average treatment effect (CATE), $\tau(x_j) = E[\text{FAMINES}_j | X = x_j]$ where $X \in \{\text{ETHNICITY}, \text{ENTITLEMENTS}\}$. While a parametric model test a specific parametric interaction, this ML model searches over the join conditional distribution of $p(\text{FAMINES}_j | X = x_j)$ for group-specific causal effects. These groups are defined by $X = x_j$.

In the second combination, scholars apply ML in the service of commonly used causal methods. Even if the statistical model of interest is a parametric model where $\hat{\beta}$ imprints the causal effect of interest, ML can service this model in an initial estimation step. While many parametric approaches that rely on two or more estimation steps can benefit from such a service, instrumental variables methods (Belloni et al. 2018; Belloni, Chernozhukov, and Hansen 2014; Carrasco 2012) and propensity score models (Alaa, Weisz, and van der Schaar 2017; Ju et al. 2017; Wyss et al. 2017) have benefited most. An instrumental variable method is an identification and estimation technique used in a situation when the targeted causal relationship between the treatment $W = w$ and an outcome, $Y = y$, is confounded by an unobserved event, $C = c$. In this situation, an instrument $Z = z$ disentangles the variation of $C$ on $Y$ from the variation of $W$ on $Y$, thereby capturing the targeted causal effect. A key assumption is that the instrument affects only the treatment, and that this affect mimic a random coin flip (Morgan and Winship 2014). Regressing the treatment on the instrument, a statistical model cleanses the variation of $w$ from the contaminating effect of the confounder $c$. The instrumental variable method comes in different versions. For example, a two-stage instrumental variable version consist of the following stages: the first stage estimates $w_i = c_0 + \gamma z_i + e_i$ where $z_i$ is an instrument and the second stage $y_i = c_1 + \beta \hat{w}_i + e_i$ uses the predicted (cleaned) version of $\hat{w}$ instead of the original $w$ (contaminated). As $z$ mimics a coin flip, $\hat{w}$ transforms the treatment assignment to an as-if random event. As the first stage is a prediction problem, recent literature developed a variety of different extension from deep-instrumental variable approaches to sample-splitting procedures strengthening the capabilities of this framework (Belloni et al. 2018; Hartford et al. 2016).

Similarly, instead of relying on logistic regression to estimate a propensity score model and running the risk of overfitting, new methods utilize AMC-type of procedures and algorithms to estimate propensity scores (Lee, Lessler, and Stuart 2010). Then, these scores are used for downstream causal estimation methods, such as inverse probability weighting. Many current methods combine the best of two worlds by predicting the treatment propensity and evaluating an outcome (regression) model (Chernozhukov et al. 2018; Laan and Rubin 2006; Nie and Wager 2018; Schuler and Rose 2017; Sverdrup et al. 2020).

In the third combination, HMC scholars use ML for policy optimization in sequential decision-making. Many problems in a dynamic-decision-making regime have a sequential structure (Russo et al. 2018). By *dynamic*, we mean a situation where a treatment assigned at time $t$ has a causal effect not only on the outcome at time $t$, but also on subsequent treatment decisions $t + k$ and covariates (Hernan and Robins 2020). For example, a doctor treating a cancer patient seeks to identify the optimal treatment with the least amount of paint (Gottesman et al. 2019, 2019; Murphy 2003). But what this doctor decides to inject into the bloodstream of this patient at time $t$ will affect the patient's pain level in the next sequence and what the doctor can switch to in future sequences if the first trial was unsuccessful. Consequently, this sort of sequential decision-making problem translates to finding the optimal policy with the desired causal effect with as few treatment steps as possible (Håkansson et al. 2020).

On the one hand, this problem of "searching-over-potential treatments (or actions, policies) to find the optimal effect" aligns with DMC ambitions of identifying a causal effect. On the other hand, this problem also has a predictive structure similar to those AMC problems of algorithms playing computer or board games (Russo et al. 2018). For example, following AMC, DeepMind's reinforcement-learning algorithms AlphaStar and AlphaGo select the optimal decision at time $t$ that predicts the best chance of eventually winning the game. These algorithms are purely predictive, with no causal component. In HMC, reinforcement-learning algorithms combine both predictive and causal components: they estimate the causal effect for this particular decision, and they predict the final outcome for a sequence of decisions. They achieve this complex task by using similar principles from the first way ML supports causal inference. Reinforcement algorithms impute potential outcomes for many possible sequences and then based on these synthetic data, they select optimal decisions. Scholars have also combined these algorithms with wearables and sensing technologies, thereby embedding sequential medical interventions directly into patients' daily lives (Liao, Klasnja, and Murphy 2020). Other scholars explore how reinforcement algorithms can be used to find optimal economic policies (Kasy 2018; Zheng et al. 2020).

In the fourth combination, scholars apply ML for causal discovery. Many new and complex scientific domains lack robust theories about how events in that domain are causally connected. In these domains, scholars lack precise DAG representations of the causal system. To remedy this lack and to fuel causal theorizing, causal-discovery algorithms suggest such DAGs (Jaber, Zhang, and Bareinboim 2018; Peters et al. 2017). Assuming the existence of data that represents all key variables of a causal system, these algorithms search over the covariate space to find and suggest DAGs inductively (Spirtes, Glymour, and Scheines 2001).

The causal-inference toolbox offers different algorithms for suggesting DAGs (Glymour et al. 2019). For example, independence-based algorithms test for permutations of conditional and unconditional independence in the joint distribution of the data $p(X_1, X_2, ..., X_k)$. For two variables, if the algorithm finds that two variables are independent $X_1 \perp X_2$, then it assigns a higher probability that these two variables are less likely causally connected. Because they are independent, their joint distribution factorizes into $p(X_1, X_2) = p(X_1)p(X_2)$, and the algorithm abstains from drawing an edge between these two variables. For three variables, if the algorithm finds $X_1 \perp X_2 | X_3$, then that means that $X_1$ and $X_2$ are independent conditional on $X_3$. This conditional independence yields three mutually exclusive interpretations. First, this finding suggests that $X_3$ is a mediator as it lies on the path between $X_1$ and $X_2$, like the following $X_1 \rightarrow X_3 \rightarrow X_2$. The joint distribution of this path factorizes into $p(X_1, X_2, X_3) = p(X_1)p(X_3|X_1)p(X_2|X_3)$. Second, $X_3$ is a mediator in the other direction $X_1 \leftarrow X_3 \leftarrow X_2$ with a joint distribution of $p(X_1, X_2, X_3) = p(X_2)p(X_3|X_2)p(X_1|X_3)$. Third, this conditional independence suggest also that $X_3$ is a common cause, $X_1 \leftarrow X_3 \rightarrow X_2$ with a joint distribution that factorizes into $p(X_1, X_2, X_3) = p(X_3)p(X_1|X_3)p(X_2|X_3)$. The algorithm needs more information to discern among these three interpretations. If the dataset has additional variables, the algorithm continues adjusting the edges and filtering the most likely DAGs from the last plausible based on other independencies.

No matter the data matrix's width or length, causal discovery suffers from at least three limitations. First, as many discovery methods rely on conditional-independence tests, perturbations in the data—either through sampling or measurement error—may flip the causality direction, especially if the true causal effect is small (Shah and Peters 2020). While conditional-independence tests are relevant for many statistical practices (Dawid 1979), such tests the foundation for many discovery algorithms (Shah and Peters 2020). Thus, if these tests are unreliable, the algorithm's results are untrustworthy. Second, causal discovery assumes that all relevant variables are measured (Robins and Wasserman 1999). That is a strong assumption because that requires that all DAGs of competing theories $T_k$ are representable in data. That requires the measurement of a vast number of variables, and that requirement will likely not be fulfilled for or even modestly complex causal systems.

Third, even if all variables were measured and there were no issues in the conditional-independence tests, causal-discovery algorithms rarely find one optimal DAG but many candidate DAGs representing the same causal system (Peters et al. 2017). The observed data can only do so much. Nonetheless, combined with domain knowledge or randomized-control trials, scholars continue the filtering of plausible and implausible DAGs. This combination of machine-suggested DAGs and human-domain knowledge of the causal system have proven fruitful in, for example, genetics. Because of the complexity of how genes interact and regulate each other, scholars are yet to determine precisely the causal direction of any genetic system.

Causal-discovery algorithms have proven useful in suggesting a causal representation of how genes regulate each other (Glymour et al. 2019). While causal discovery is a vibrant field of research with promising capacity, other research fields are yet to test their usefulness.

In sum, through the four ML-powered ways to causal inference, DMC and AMC practices merge into a new culture with modified statistical practices. This new culture, HMC, weaves prediction and inference into synthesized procedures captured by $\hat{\tau}$, yielding the traditional distinction of $\hat{Y}$ and $\hat{\beta}$ superfluous.

## 3.2 ML for data acquisition

That scholars collect all the necessary variables representing the events of a causal system of interest is a prerequisite for causal inference. The primary and minimum variables for a causal evaluation in observational studies are a treatment $W$ (a policy, action, or exposure), an outcome $Y$, and a confounder $C$. Figure 3 shows a DAG containing the basic set of variables for causal analysis in observational settings. If either the treatment or outcome is unobserved, quantifying their causal connection $\tau$ is impossible; if only the confounder is unobserved, statistical models will produce biased results of $\tau$. Confounders are variables that affect both the treatment and the outcome. The magnitude of this bias depends on how strongly a confounder affects both these variables (Hernan and Robins 2020). For any scientific domain, collecting high-quality data about the causal system of interest remains, therefore, a crucial task.

[Figure 3 about here]

The second practice of HMC mobilizes AMC-type algorithms to measure data. Table 2 defines the key characteristics of these practices. While in DMC-type practices, scholars rely more on analog methods to measure data and less on digital ones, in AMC, this emphasis is reversed. An *analog-measurement method* relies predominantly on humans to measure and structure information about causal events. Conducting surveys, interviews, laboratory experiments, and ethnographic studies are examples of analog methods. A *digital-measurement method* is any method that relies mainly on algorithms to extract data from structured or unstructured computerized sources. A *computerized* source is a piece of information existing as "1s" and "0s" on a computer. By *structured*, we mean information that exists as a tidy data matrix in well-defined variables and values. Conversely, *unstructured* information has yet to be preprocessed for a meaningful structure. Historical digitized archival documents are one example of unstructured information (Salganik 2017). Processing a large amount of structured and unstructured information is one of the tasks where ML excels. One the one hand, ML learns to find clusters of similar information in unstructured data; on the other hand, it finds patterns in large data to predict an outcome.

For example, mobilizing research assistants to code up the political content of arhival policy documents an analog method (Daoud, Reinsberg, et al. 2019); training natural language processing (NLP) algorithms to do the same thing is a digital method (Grimmer and Stewart 2013). Likewise, in famine research, employing assistants to analyze geographical maps to code events of drought is an analog method; training image-recognition algorithm to detect drought in satellite images is a digital method (Mahecha et al. 2020). Each method has its strengths and weaknesses. As analog methods rely on humans, these methods are better tuned to measuring sensitive content, but they are slower and more costly. While digital methods still require human supervision for training data and interpreting the content of unsupervised results, they are still faster and cheaper (Salganik 2017).

In the digital era, several additional petabytes of data are made available every year. Equipped with DMC-logic for sampling issues, HMC scholars use AMC practices to efficiently sift through these data to measure the causal system of interest. Although the role of digital methods is considerable in HMC, analog methods remain an essential part of HMC practices. For example, unsupervised algorithms reduce high-dimensional data (e.g., an archival document) to low-dimensional representations (e.g., a topic-model distribution), but these low-dimensional representations still require that scholars interpret the meaning of these representations. Unsupervised ML is algorithms that reduce dimensionality in a set of covariates $X$, without any reference to a specific outcome, $Y$. Similarly, supervised algorithms will still require high-quality training data prepared by scholars. Supervised ML is algorithms that are tailored to predict a specific outcome, $Y$, using $X$. Often these data rely on surveys or qualitative coding of digital sources, based on the scholars' expertise about the causal system.

When applying ML to synthesize variables from digital sources, scholars conduct a form of data measurement. For example, instead of asking people directly about their material living standards, scholars let a machine capture these standards from digital sources (Jean et al. 2016). However, a limitation of using ML to conduct such measurement is the added error in the reconstruction of the variable of interest.

The total-survey-error approach helps characterize the sources of error in traditional surveying (Groves and Lyberg 2010). Analog surveys pose specific challenges arising from formulating questionnaires, conducting interviews, and other disturbances affecting measurements. This approach provides a framework to characterize what these sources of error are when a scholar measures an event $X^*$ and quantifies it as a variable in the data $X'$. Although different errors exist, they categorize either as bias (systematic error) or variance (random error). *Systematic error*, $e_s$, is any reoccurring information that shifts the sample estimate $X'$ away from the true value $X^*$ in a consistent manner. Inaccurately calibrated instruments usually cause such shifts. An instrument is a means to acquire information. For example, poorly phrased wordings in a questionnaire (the instrument) will lead to over- or underreporting about a respondent's behavior. Systematic error can only be reduced by improved calibration of the instrument. *Random error*, $e_r$, is the natural variation arising from sampling procedures that affect the accuracy of the instrument. Random errors cancel each other out as sample size increases because negative deviations for one individual $X'_i$ eventually, cancel out by a positive deviation for another $X'_j$. Putting these two sources of error together, a survey will always be an imperfect representation of any causal system, as described by the following formula, $X^* = X' + e_s + e_r$.

While scholars can improve their instruments and surveying execution to reduce these two sources of error, they will eventually hit a limit where they start trading bias for variance or vice versa. Both analog and digital methods suffer from the same limitations, yet digital methods inject additional errors (Salganik 2017).

As many supervised ML algorithms rely on the analog data source for training samples, these algorithms can only recreate imperfect representations of $Y'$ and not of the true event $Y^*$. This imperfection arises from training and testing the algorithms (Hastie et al. 2009). For example, in famine and poverty research, many scholars use analog methods to measure people's living condition, $Y^*$, by surveying people's income or material assets, $Y'$ (Corsi et al. 2012). In designing and executing their surveys, scholars encounter both systematic and random errors, $e_1 = e_{1s} + e_{1r}$, that will distort their sample estimate, $Y'$. Nonetheless, $Y'$ is the best they can produce to represent $Y^*$. Subsequently, another group of scholars aims to speed up the surveying of poverty by relying on a digital source such as satellite images (Blumenstock, Cadamuro, and On 2015; Jean et al. 2016; Yeh et al. 2020). These images reveal the living conditions of people. So, this group collects $Y'$ as a training sample $y'_1, y'_2, \ldots, y'_n$ for their ML algorithm, and digitally available satellite archives, $S$. Their goal is to train an algorithm $f$ to measure (predict) income $\hat{Y}'$ from the pixel features of these satellite images $S$. This procedure constitutes defining $Y' = f(S) + e_2$, where $f(S) = \hat{Y}'$ and an error composed of $e_2 = e_{2s} + e_{2r}$. Although this ML-satellite approach to measure poverty is faster than letting humans surveying poverty, each new ML step added to represent $\hat{Y}'$ continues eroding at the human-survey produced $Y'$. This $Y'$ is already an imperfection of $Y^*$. Consequently, each ML step induced a new error resulting in the following mathematical relationships,

$$Y^* = Y' + (e_{1s} + e_{1r})$$
$$Y^* = f(S) + (e_{2s} + e_{2r}) + (e_{1s} + e_{1r})$$
$$Y^* = \hat{Y}' + (e_{2s} + e_{2r}) + (e_{1s} + e_{1r})$$
$$Y^* - (e_{2s} + e_{2r}) - (e_{1s} + e_{1r}) = \hat{Y}'$$

Systematic and random errors compound from the analog (i.e., $e_{1s} + e_{1r}$) and digital methods (i.e., $e_{2s} + e_{2r}$). The more algorithmic transformations added on top of the first procedure $\hat{Y}'$, the more these errors are likely to propagate, continuing to drifting away from $Y^*$.

Although measuring events digitally induces an additional error, digital approaches are usually faster and less costly. These two advantages have prompted HMC-influenced scholars to new sources of data to impute the missing data to populate their representation of the causal system of interest—a procedure that Bareinboim and Pearl (2016) formalized under the name *data-fusion problems*. For example, scholars use topic models or other representations to summarize text to capture confounding (Åkerström, Daoud, and Johansson 2019; Blei 2012; Blei, Ng, and Jordan 2003; Daoud and Kohl 2016; Egami et al. 2018; Mozer et al. 2020; Roberts, Stewart, and Airoldi 2016), images to measure outcomes (Jean et al. 2016), a corpus of health record to record patients' background (Hsu et al. 2020), and video and audio for other representations (Hwang, Imai, and Tarr 2019; Knox and Lucas 2019). These new measures are then used for downstream causal inference tasks. Synthesizing analog and digital approaches will remain a crucial part of the scientific endeavor.

Usually, scholars use ML to predict events with the hope of lowering measurement errors as much as possible. Nevertheless, to satisfy a societal or political value, occasionally, errors are injected deliberately and algorithmically. For example, to protect individuals' privacy, the U.S. Census Bureau is under the obligation to ensure that confidential information remains private while producing useful demographic information. Since 1840, the Bureau has used various manual mechanisms to mask private information. Such manual mechanisms often pertained to aggregating information (e.g., age, ethnicity, or race) to local areas or regions. However, during the digital era, the risk of malicious-information hacking has risen dramatically. Such hackers want as precise and private individual-level information as possible to launch targeted attacks. The *database-reconstruction theorem* shows that hackers can use algorithms to reconstruct individual-level data with high precision (Dinur and Nissim 2003). *Differential privacy* is the study of such algorithms and the tradeoff between privacy and usability of statistical databases. From a statistical and causal inference perspective, a critical scientific task is then to find an optimal tradeoff between protecting privacy while enabling scholars to use data with as much precision as possible (Evans and King 2020; Gong and Meng 2020). In other words, while ML for data acquisition will often pertain to reducing measurement error, in some cases, it will be about balancing privacy and inference.

## 3.3 ML for theory prediction

A vital goal of the scientific endeavor is to explain not only past observations of how an event causes another, $W \rightarrow Y$, but also predict future instances of these causal events (Gelman and Imbens 2013). This goal translates to testing a theory's predictive power: *theory prediction*, for short (Kleinberg, Liang, and Mullainathan 2017; Peysakhovich and Naecker 2017; Watts et al. 2018, 2018). Given that a scientific theory, $T_k$, and its corresponding DAG, $G_k$, already received statistical support in data—meaning that $W$ affects $Y$ by an amount of $\hat{\tau}$—then another test of the usefulness of a scientific theory is to evaluate how predictive $T_k$ is for other populations (Billheimer 2019). While Section 3.1 "ML for causal inference" describes how HMC scholars test, discover and build theories on causal relationship, this Section shows how these scholars evaluate the predictive performance of their theories. Table 2 shows the key characteristics of these three HMC practices.

Despite that Breiman argues in favor of prediction as a tool for fueling the scientific endeavor, he remains unclear about how exactly prediction supports causal inference. We define a *DMC-prediction* as a recreation of an event $Y$ that is as similar as possible to the true event yet not observed but generated by a validated statistical model specified under $T_k$ and its DAG, $G_k$. In our famine example, the statistical model $f_1(\text{FOODSCARCITY}_i)$ representing a Malthusian DAG $G_1$ constitute one such validated model. A DMC-prediction of famines is then $\hat{Y}_{G_1} = f_1(\text{FOODSCARCITY}_i)$, for famine events $Y_i$ with characteristics $\text{FOODSCARCITY}_i$, for cases $i$ that the statistical model $f_1$ has not been observed before (Conklin et al. 2018; Daoud 2007, 2010, 2011, 2015b; Daoud and Nandy 2019; Nandy, Daoud, and Gordon 2016). In contrast, an *AMC-prediction* is a recreation of an event that is as similar as possible to the true event yet not observed, but it is not necessarily conditioned on any validated model of $T_k$. In its distilled form, an AMC-prediction is a pure prediction problem, using any algorithm and data source (Kleinberg et al. 2015). For example, an AMC-prediction of our famine-example involves collecting viable input variables that carry any association to the outcome (famines), training any ML model, and then evaluating this model's predictive power on a held-out set (Okori and Obua 2011). AMC-predictions constitute a horserace among candidate algorithms competing for the best predictive performance—Kaggle-style competitions.[7] DMC-predictions are also horseraces but among scientific theories and their statistical representations (Daoud, Kim, and Subramanian 2019).

HMC-predictions are DMC-predictions, but as HMC-predictions rely on ML, it submits its prediction practices to the principles dictated by AMC. As discussed previously, two essential principles are regularizations and evaluation using held-out samples. These two principles minimize overfitting that likely arises when scholars attempt to squeeze in more variables into their model or tweak their models' functional form to fit the data better in-sample. Additionally, as future observations are unobserved in the present, held-out samples stand-in for these missing observations (Watts 2014). Highly predictive theories have a small difference, $\epsilon_{\hat{Y}_{G_k}}$, between what the theory says will happen $\hat{Y}_{G_k}$ and what eventually did happen, $Y$, such that $\epsilon_{\hat{Y}_{G_k}} \approx Y - \hat{Y}_{G_k}$. Two theories competing for superiority in explaining $W \rightarrow Y$, generate their respective predictions on new data, and then evaluate their predictive performance. They do so by using their respective DAGs, $G_k$, which defines how context $X$ affects $W \to Y$. The theory scoring the smallest $\epsilon_{G_k}$, wins a theory-prediction contest.

---

[7] www.kaggle.com

However, several substantive theories abstain from defining all the structural relationships in the causal system because of the complexity of many causal systems. Defining all structural relationships means characterizing not only the relationship of interest, that is between the cause and effect, $W \to Y$, but also characterizing how the context variables, $X_1, X_2, \ldots, X_z$ are related amongst each other and to $W \to Y$. For example, a theory defining exactly how the social environment interacts with human genes to explain students' university grades is complex because scientists do not know enough about these causal relationships (Beauchamp 2016; Courtiol, Tropf, and Mills 2016). Thus, evaluating how well such a theory predicts university grades may be a difficult task, resulting in poor predictions $\hat{Y}_{G_{gene}}$.

A second way to test a theory's predictive power is to evaluate whether the identified causal effect $\hat{\tau}_{G_k}$, using DAG $G_k$, found in one population can be verified in another. In computer science, these verifications are called *transportability* of results (Pearl and Bareinboim 2014), and they are also known as *transfer learning*, *life-long learning*, and *domain adaptation* (Chen and Liu 2018; Johansson, Sontag, and Ranganath 2019); in statistics, and elsewhere, they go under the name of *generalizability* or *external validity* (Deaton and Cartwright 2016). For predicting future causality, some scholars call these verifications *forward-causal questions* (Gelman and Imbens 2013). For example, finding that a medical treatment works for patients in one hospital, scholars may ask how well this treatment generalizes for other hospitals; similarly, how well do social policies from one country (e.g., Sweden) transport to other countries. The more populations for which $\hat{\tau}$ exists and has a similar value, the more support a theory receives.

To systematically evaluate the predictive performance of theories' $\hat{Y}_{G_k}$ and $\hat{\tau}_{G_k}$, require a *common task framework* (Donoho 2017). This framework is a set of principles defining how this predictive performance is scored and evaluated. Its purpose is to ensure the comparability of results. Such a framework has at least three principles. First, scholars require a publicly available training dataset for which they may operationalize their respective scientific theories $T_k$, articulate their causal assumptions $G_k$, and formulate their statistical models $f_{G_k}$. This dataset must be sufficiently rich to accommodate many different plausible theories $T_k$. Second, as in any competition, scholars have to make themselves known to each other and agree to the competition's rules. These rules must at least specify one estimand (quantity of interest) and how scholars' estimators will be evaluated (e.g., minimizing mean squared error). Third, the competing scholars have to designate a scoring referee for which they can submit their estimators. This referee automatically and objectively checks each scholar's estimator against a held-out dataset that has been kept secure behind a firewall during the competition. The referee presents the results and estimators transparently, enabling the competing scholars to reproduce each other's results, thereby enhancing scientific learning.

Much of the success of AMC is based on well-functioning common task frameworks (Donoho 2017). Often there are clear outcomes $Y$ defined, and transparent procedures for scoring each competitor prediction, $\hat{Y}$. For example, much of the development of image-recognition algorithms owns its success to the publicly available data source, ImageNet (Deng et al. 2009). It gave deep-learning scholars a common benchmark, which new algorithms could be tested against. In DMC, setting up a similar infrastructure is challenging because the estimand is often a causal effect, $\tau$, which is per definition, unobserved in the data. As neither the scoring referee nor the scholars know the true $\tau$, there is no way to score contending estimators. This implies that a common task framework for causal problems has to be complemented with additional assumptions for the competition to work. One way of handling this insufficiency is to combine observational data with randomized-control trial data (Lin et al. 2019). As RCTs require the least assumptions, one target estimand is the average treatment effect in RCT, $\tau_{RCT}$. However, this target will work if the RCT has high internal and external validity. Another way of handling this insufficiency is to rely on simulations in which the causal effect is know exactly, $\tau_{sim}$. Although that solves the problem of the target estimand, it introduces the problem that the data is merely an artificial representation of a causal system.

While establishing a common task framework to evaluate causality remains a challenge in many disciplines—especially in the social sciences—several common task frameworks focus on an observable estimand: predicting outcomes $Y$. For example, the Fragile Families Challenge is a scholarly mass collaboration tailored to predict six life outcomes for children age 15 (Salganik et al. 2019; Salganik, Lundberg, et al. 2020; Salganik, Maffeo, and Rudin 2020). These outcomes are child grade point average (GPA), child grit, household eviction, household material hardship, caregiver layoff, and caregiver participation in job training. This Challenge received 437 scholarly competitors (of which some worked in teams) that resulted in 160 valid submissions. All these submissions were evaluated using *means squared error* in held-out data. In the social sciences, this challenge is among the first to devise a common task framework for the advancement of science.

## 4  Discussion and conclusion

At the beginning of the twenty-first century, Leo Breiman described two cultures of statistical modeling and their influence on statistical practices. While the data modeling culture (DMC) is the *modus operandi* in applied research, the algorithmic modeling culture (AMC) dominates the industry, engineering, and policy practices. Given the new scientific opportunities and challenges arising in the digital era, scholars have developed various techniques fueling the scientific endeavor. In these developments, a new modeling culture has evolved. As this culture mutated from DMC and AMC, we named it the hybrid modeling culture (HMC).

This article has identified the main characteristics of HMC and showed how it synthesizes components of DMC and AMC under the umbrella of causal inference. First, the overarching aim of HMC is to further the production of scientific knowledge. HMC copies this aim from DMC, where the overarching goal is to explain how *X* causally affects *Y*. Scholars achieve this goal by assuming a causal system, $G_{T_k}$ stipulated by a substantive theory $T_k$. Under these assumptions, they test competing explanations against data, refute, revise, or update theories depending on these tests' results. Second, HMC does not restrict itself to the commonly used statistical models offered by DMC for statistical and causal inference but incorporates the range of intelligent algorithms offered by AMC. As scholars combine AMC-type algorithms for DMC-inspired inference, they erode the traditional distinction between prediction $\hat{Y}$ and statistical inference $\hat{\beta}$. This erosion recasts the scientific problem into $\hat{\tau}$ problems—identifying ways to impute (predict) potential outcomes $Y^1$ and $Y^0$.

We conclude that instead of upholding the dichotomy between AMC-prediction and DMC-inference, the scientific endeavor gains more by embarrassing both synthetically. HMC provides a way to think about how this synthesis is possible in combination with data science (Meng 2020) while still maintaining the scientific endeavor's higher goal: explaining reality.

Malthus, Thomas Robert. 1826. *An Essay on the Principle of Population, or a View of Its Past and Present Effects on Human Happiness; with an Inquiry into Our Prospects Respecting the Future Removal or Mitigation of the Evils Which It Occasions*. 6th ed. London: John Murray.

Marathe, Madhav, and Anil Kumar S. Vullikanti. 2013. "Computational Epidemiology." *Communications of the ACM* 56(7):88–96. doi: 10.1145/2483852.2483871.

Meadows, Donella H., Dennis L. Meadows, Jørgen Randers, and William W. Behrens III. 1972. *The Limits to Growth : A Report for The Club of Rome's Project on the Predicament of Mankind*. London: Earth Island.

Meng, Xiao-Li. 2020. "What Is Your List of 10 Challenges in Data Science?" *Harvard Data Science Review*. doi: 10.1162/99608f92.a3e88876.

Merton, Robert K. 1968. *Social Theory and Social Structure*. New York: The Free Press.

Molina, Mario, and Filiz Garip. 2019. "Machine Learning for Sociology." *Annual Review of Sociology* 45(1):27–45. doi: 10.1146/annurev-soc-073117-041106.

Morgan, Stephen L., and Christopher Winship. 2014. *Counterfactuals and Causal Inference: Methods and Principles for Social Research*. 2 edition. New York, NY: Cambridge University Press.

Mozer, Reagan, Luke Miratrix, Aaron Russell Kaufman, and L. Jason Anastasopoulos. 2020. "Matching with Text Data: An Experimental Evaluation of Methods for Matching Documents and of Measuring Match Quality." *Political Analysis* 28(4):445–68. doi: 10.1017/pan.2020.1.

Mullainathan, Sendhil, and Jann Spiess. 2017. "Machine Learning: An Applied Econometric Approach." *Journal of Economic Perspectives* 31(2):87–106. doi: 10.1257/jep.31.2.87.

Murphy, S. A. 2003. "Optimal Dynamic Treatment Regimes." *Journal of the Royal Statistical Society: Series B (Statistical Methodology)* 65(2):331–55. doi: 10.1111/1467-9868.00389.

Nandy, Shailen, Adel Daoud, and David Gordon. 2016. "Examining the Changing Profile of Undernutrition in the Context of Food Price Rises and Greater Inequality." *Social Science & Medicine* 149:153–63. doi: 10.1016/j.socscimed.2015.11.036.

Nie, Xinkun, and Stefan Wager. 2017. "Learning Objectives for Treatment Effect Estimation." *ArXiv:1712.04912 [Econ, Math, Stat]*.

Nie, Xinkun, and Stefan Wager. 2018. "Quasi-Oracle Estimation of Heterogeneous Treatment Effects." *ArXiv:1712.04912 [Econ, Math, Stat]*.

Noble, Denis. 2002. "The Rise of Computational Biology." *Nature Reviews Molecular Cell Biology* 3(6):459–63. doi: 10.1038/nrm810.

Okori, Washington, and Joseph Obua. 2011. "Machine Learning Classification Technique for Famine Prediction." 6.

# 6 Tables

*Table 1: Central practices of two statistical cultures*

|  | **Data-modeling culture (DMC)** | **Algorithmic-modeling culture (AMC)** |
|---|---|---|
| **Exemplifying question** | What is the causal relationship between food supply and famines? | How well can famines be predicted from available data? |
| **Goal** | Estimating unbaised parameters for causal estimation, to populate the magnitudes of the edges of a DAG. | To develop and train an algorithm $f$ for accurate prediction. |
| **A key assumption** | Assuming a DAG, a stipulated and interpretable statistical model such as $y_i = c_0 + \beta w_i + e_i$ produces unbiased estimates of the true causal quantity $\beta$. | The algorithm $f$ can produce accurate predictions of $Y$ from data source, $D$. |
| **Limitation** | Although the parametric model is interpretable, its statistical structure may be a poor representation of the causal system. | Although $f$ produces accurate predictions, the model is a black-box restricting causal interpretations. |
| **Quantity of interest** | $\hat{\beta}$ | $\hat{Y}$ |

Notes: a) in the equation $y_i = c_0 + \beta w_i + e_i$, the outcome is $y_i$ and the treatment is $w_i$. The variable $c_0$ is the intercept and $e_i$ is the residual.

*Table 2: Central practices of the hybrid-modeling culture (HMC)*

|  | **ML for causal inference** | **ML for data acquisition** | **ML for theory prediction** |
|---|---|---|---|
| **Exemplifying question** | What is the causal relationship between food supply and famines? | Can food availability be measured from satellite images? | How well does the Malthusian theory of famines predict new famines? How does it compare to a Senian theory? |
| **Goal** | Imputing potential outcomes for causal estimation, to populate the magnitudes of the edges of a DAG. | Producing new indicators from digital sources, $D$, to populate the nodes of a DAG. | Comparing the predictive power of two or more theories' DAGs, $\hat{Y}_{G_1}, \hat{Y}_{G_2}, \dots, \hat{Y}_{G_k}$, for new realizations of an outcome, $Y$. |
| **A key assumption** | The algorithm $f$ produces unbiased estimates of the true causal quantity $\tau$, assuming a DAG. | The algorithm $f$ can measure the true quantity of the variable of interest $(X, W, Y)$ from a digital source, $D$. | The algorithm $f$ is an appropriate representation of $G_k$ to predict, $Y$. |
| **Quantity of interest** | $\hat{\tau} = \hat{Y}_i^1 - \hat{Y}_i^0$ | $\hat{X}, \hat{W}, \hat{Y}$ | $\epsilon_{\hat{Y}_{G_k}} \approx Y - \hat{Y}_{G_k}$ or $\epsilon_{\hat{\tau}_{G_k}} \approx \tau - \hat{\tau}_{G_k}$ |

*Table 3: A toy dataset illustrating the fundamental problem of causal inference*

|      | $Y$ | $Y^1$ | $Y^0$ | $W$ | $\tau$ | $X$ |
|------|-----|-------|-------|-----|--------|-----|
| **Jane** | 20 | 20 | ?  | 1 | ? | 10 |
| **John** | 30 | 30 | ?  | 1 | ? | 11 |
| **Joe**  | 25 | ?  | 25 | 0 | ? | 10 |
| **Jan**  | 22 | ?  | 22 | 0 | ? | 11 |

# 7  Figures

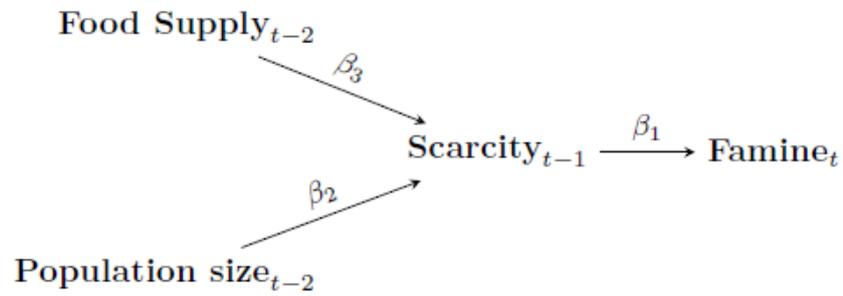

(a) Malthus' theory of famines

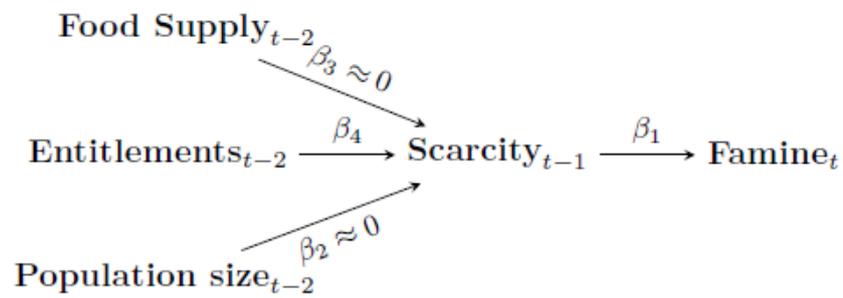

(b) Sen's theory of famines

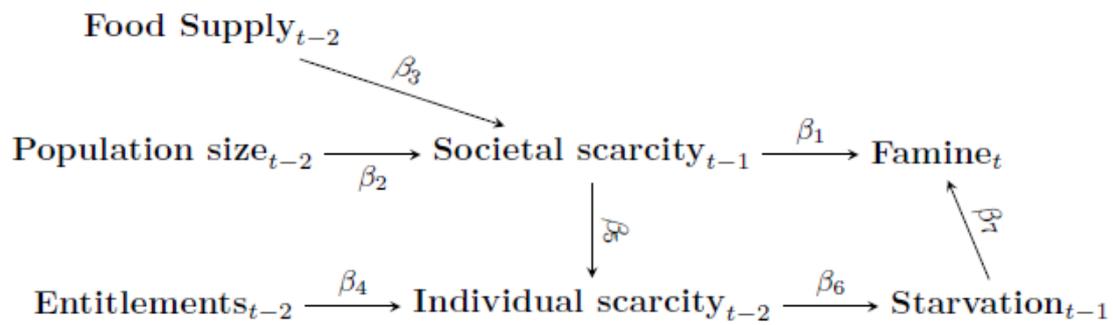

(c) A synthesized theory of famines

*Figure 1: Directed acyclic graphs depicting stylized causal systems of famine*

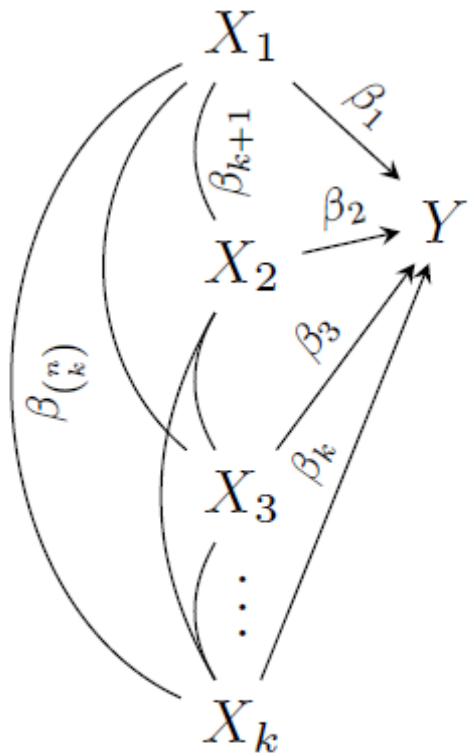

(a) A covariate set associated with an outcome

Figure 2: Bayesian network in the Algorithmic modeling culture

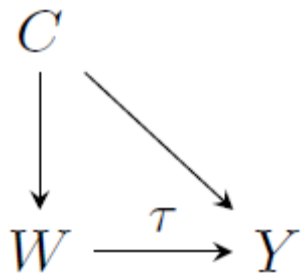

(a) Basic components of a causal system

Figure 3: The basic set of variables for causal analysis in observational settings